\documentclass[12pt,letterpaper,notitlepage]{article}
\usepackage[margin=1in]{geometry}
\usepackage{graphicx}
\usepackage{amssymb}
\usepackage{amsmath}
\usepackage{amsbsy}
\usepackage{epstopdf}
\usepackage{pdflscape}
\usepackage{setspace}
\usepackage{nicefrac}
\usepackage{multirow}
\usepackage{booktabs}
\usepackage{natbib}
\usepackage{enumerate}
\usepackage{makeidx}  %%% standard INDEX
\usepackage[titletoc]{appendix}
\usepackage{fancyhdr}

\usepackage{longtable}

\usepackage[latin1]{inputenc}
\usepackage{geometry}
\usepackage{graphicx}
\usepackage{epstopdf}
\usepackage{amsmath}
\usepackage{amsfonts}
\usepackage{amssymb}
\usepackage[parfill]{parskip}	% activate to begin paragraphs without an indent
\usepackage{color}
\usepackage{bm}
\usepackage{float}
\usepackage{setspace}		% 1.5 spacing
\onehalfspace
\usepackage{hyperref}
\usepackage{subfigure}

\usepackage[font=small,labelfont=bf]{caption} %activate for figure and table captions to be put in smaller font

\onehalfspace
\allowdisplaybreaks

\begin{document}

\title{Recent Advances on Estimating Population Size with Link-Tracing Sampling}

\author{\begin{large}Kyle Vincent\end{large}\footnote{ \textit{email}: kyle.shane.vincent@gmail.com}}

\date{\today}

\maketitle

\begin{abstract}
\noindent
A new approach to estimate population size based on a stratified link-tracing sampling design is presented. The method extends on the \cite{Frank1994} approach by allowing for heterogeneity in the initial sample selection procedure. Rao-Blackwell estimators and corresponding resampling approximations similar to that detailed in \cite{Vincent2017} are explored. An empirical application is provided for a hard-to-reach networked population. The results demonstrate that the approach has much potential for application to such populations. Supplementary materials for this article are available online.

\end{abstract}

%\thispagestyle{empty}

%\pagenumbering{gobble}

\pagenumbering{arabic}

\clearpage

\section{Introduction}

There is a growing demand for practical methods to estimate population size and other quantities of hidden networked populations. A new and flexible approach that extends on previous work is presented.

The approach is applied to an empirical population at risk for HIV/AIDS. The design commences with the selection of an initial sample where selection probabilities depend on strata memberships. Links are traced from sampled members with probabilities also dependent on strata memberships. Consistent estimation for population quantities is made with a design-based approach to inference. Preliminary estimators are based on information in the initial sample. Improved estimators are obtained via the Rao-Blackwell theorem, which incorporates information from units added to the sample through link-tracing.

Advantages of using the novel approach over existing methods are: 1) it allows/accounts for heterogeneity in initial sample selection probabilities; 2) it has the ability to harness nominations from conspicuous/certainty individuals, or even those external to the target population, in the inference procedure to substantially improve the precision of estimators; for example, such nominations may come from those individuals conceivably sampled with probability one (typically, the social stars of the population), a pilot study, or from researchers familiar with the target population via a prior study; 3) it allows for heterogeneity in how nominations are defined between strata; for example, links originating from a stratum of males and which are directed towards females may be based on sexual contact, and within a stratum of males may be based on sharing drugs; and 4) it bases approximations for the computationally intensive improved (Rao-Blackwellized) estimators on an updated and efficient Markov chain Monte Carlo resampling procedure that depends on a suitable convergence diagnostic test.

\section{Sampling Design}

Let $U=\{1,2,...,N\}$ be the set of units/members of the population. Suppose there are $K$ strata the population is partitioned into, possibly based on demographic configurations. In keeping with the notation of \cite{Frank1994} define $U_k$ to be those individuals in stratum $k$ for $k=1,...,K$. Define $N_k=|U_k|$ to be the size of stratum $k$. Define $y_{i,j}= 1$ if unit $i$ nominates unit $j$ and 0 otherwise, where nominations are based on predetermined relationships that may be functions of strata. Define $y_{i,i}=1$ for all $i=1,2,...,N$. Define $y_i^{k+}$ to be the number of nominations from unit $i$ to stratum $k$. Define $w$ to be the number of links in the graph, $w=\sum\limits_{i,j}y_{i,j}$, and $w_{l,k}$ to be the number of links from $U_l$ to $U_k$, $w_{l,k}=\sum\limits_{i \epsilon U_l}\sum\limits_{j \epsilon U_k}y_{i,j}$. Define $z_i$ to be the response(s) of interest attached to unit $i$. For example, this may be an indicator of drug-use status or the individual/node-degree.

The initial sample, $S_0$, is selected via a Bernoulli sampling design within each stratum; let $\alpha_k$ be the probability a unit in stratum $k$ is selected for the initial sample, and $S_{0k}=S_0\cap U_k$. Define $x_i = 1$ if unit $i$ is selected for $S_0$ and 0 otherwise. Selection for members of the first wave is carried out as follows. For any unit $i\ \epsilon\ S_{0l}$  and $j\ \epsilon\ U_k\setminus S_{0k}$ where $y_{i,j}=1$, $\beta_{l,k}$ is defined to be the probability the link is traced so that unit $j$ is added to the sample for the first wave.  Define $S_1$ to be those units selected for the first wave of the sample, and $S_{1k}=S_1\cap U_k$. For each individual $i\ \epsilon\ S=S_0\cup S_1$ define $t_i=0$ if the unit is selected for the initial sample and $t_i=1$ if the unit is selected for the first wave. The data observed upon selecting the sample is $d_0=\{i,y_i^{k+},y_{i,j},z_i,t_i: i,j\ \epsilon\ S, k=1,...,K\}$.

\section{Estimation}

\subsection{Population Size Estimation}

Define $n_{0k}=|S_{0k}|$, $r_{k,k}$ to be the number of non-self-nominated (non-loop) links within $S_{0k}$, $r_{l,k}$ to be the number of links from $S_{0l}$ to $S_{0k}$ for $l\neq k$,  $s_{k,k}$ to be the number of links from $S_{0k}$ to $U_k\setminus S_{0k}$, and $s_{l,k}$ to be the number of links from $S_{0l}$ to $U_k\setminus S_{0k}$ for $l\neq k$. The expectations of these statistics are
\begin{align}
E[n_{0k}] =& E\bigg[\sum\limits_{i \epsilon U_k} x_i\bigg] = \alpha_kN_k,\\
E[r_{k,k}] =& E\bigg[\sum\limits_{\underset{i\neq j}{i,j \epsilon U_k:}} x_i x_j y_{i,j}\bigg] = \alpha_k^2(w_{k,k}-N_k),\\
E[r_{l,k}] =& E\bigg[\sum\limits_{i \epsilon U_l}\sum\limits_{j \epsilon U_k} x_i x_j y_{i,j}\bigg] = \alpha_l\alpha_kw_{l,k},\\
E[s_{k,k}] =& E\bigg[\sum\limits_{\underset{i \neq j}{i,j \epsilon U_k:}} x_i(1-x_j) y_{i,j}\bigg] = \alpha_k(1-\alpha_k)(w_{k,k}-N_k),\ \text{and}\\
E[s_{l,k}] =& E\bigg[\sum\limits_{i \epsilon U_l}\sum\limits_{j \epsilon U_k} x_i(1-x_j)y_{i,j}\bigg] = \alpha_l(1-\alpha_k)w_{l,k}.
\end{align}
The aforementioned equations lead to the following method-of-moments estimator for $N_k$,
\begin{align}
\hat{N}_k = n_{0k}\bigg(\frac{\sum\limits_{l=1}^K r_{l,k} + \sum\limits_{l=1}^K s_{l,k}}{\sum\limits_{l=1}^K r_{l,k}}\bigg).
\label{stratum_pop_est}
\end{align}
To show the consistency of this estimator, assume that for all $k=1,...,K$, $\alpha_k \rightarrow 0$ and $N_k\rightarrow \infty$ in such a way that $\alpha_kN_k \rightarrow \infty$. Define $A_{i,k}$ to be the units in stratum $k$ nominated by unit $i$, $A_{i,k}=\{j\ \epsilon\ U_k: y_{i,j}=1\}$, and $B_{k,i}$ to be the units in stratum $k$ which nominate unit $i$, $B_{k,i}=\{j\ \epsilon\ U_k:y_{j,i}=1\}$.  Assume that for all $i\ \epsilon\ U_k$ nominations from within $U_k$ are bounded so that $|A_{i,k}|, |B_{k,i}| \leq M_{k,k}$, and for all $j\ \epsilon\ U_l$ nominations from $U_l$ to $U_k$ are bounded so that $|A_{j,k}|, |B_{l,i}| \leq M_{l,k}$. Hence, $(w_{k,k}-N_k) \leq N_kM_{k,k}$ and $(w_{l,k}-N_l) \leq N_lM_{l,k}$. As shown in the supplementary materials, these assumptions imply that
\begin{align}
Var(r_{k,k}) &\leq \alpha_k^2N_kM_{k,k} +2\alpha_k^3M_{k,k}^2= O(\alpha_k^2N_k),\\
Var(r_{l,k}) &\leq \alpha_l\alpha_kN_lM_{l,k} +\alpha_l^2\alpha_kN_lM_{l,k}^2 + \alpha_l\alpha_k^2N_lM_{l,k}^2= O(\alpha_l\alpha_kN_l),\\
Var(s_{k,k}) &\leq \alpha_k(1-\alpha_k)N_kM_{k,k}+\alpha_k^2(1-\alpha_k)N_kM_{k,k}^2 +\alpha_k(1-\alpha_k)^2N_kM_{k,k}^2=O(\alpha_kN_k),\ \text{and}\\
Var(s_{l,k}) &\leq \alpha_l(1-\alpha_k)N_lM_{l,k}+\alpha_l^2(1-\alpha_k)N_lM_{l,k}^2 +\alpha_l(1-\alpha_k)^2N_lM_{l,k}^2=O(\alpha_lN_l).
\end{align}
Together, these equations imply that $\frac{n_{0k}}{\alpha_kN_k}, \frac{\sum\limits_{l=1}^Kr_{l,k}+\sum\limits_{l=1}^Ks_{l,k}}{\alpha_k(w_{k,k}-N_k)+\sum\limits_{l \neq k}\alpha_lw_{l,k}}$, and
$\frac{\sum\limits_{l=1}^kr_{l,k}}{\alpha_k^2(w_{k,k}-N_k)+\sum\limits_{l\neq k}\alpha_l\alpha_kw_{l,k}}$ all converge in probability to 1 since their expectations are 1 and their variances tend to 0. Hence, $\hat{N}_k$ is a consistent estimator for $N_k$ and $\hat{N}=\sum\limits_{k=1}^K\hat{N}_k$ is a consistent estimator for $N$. In the simulation studies each of the stratum size estimators are stabilized in a manner that mimics the bias-adjusted Lincoln-Petersen estimator \citep{Chapman1951}; a value of one is added to each of $n_{0k}, r_{k,k}, r_{l,k}, s_{k,k}, s_{l,k}$ and the corresponding sum is subtracted from the estimator.

Of considerable note is that utilizing the statistics based on nominations originating from all strata results in a strata size estimator with a faster rate of consistency than that based on nominations originating solely from within the stratum. Hence, one can expect less-bias with the estimator based on the stratified setup.

One argument for why the estimator in Expression \ref{stratum_pop_est} works as a stratum size estimator is given as follows. In a two-sample mark-recapture study the Lincoln-Petersen estimator is the typical choice for an estimator of the population size. To work as a population size estimator only one sample need be selected completely at random while the other can correspond with a ``fixed-list". In the setup presented in this paper the Bernoulli initial sample corresponds with the sample selected completely at random and nominated individuals correspond with the ``fixed-list".

\cite{Frank1994} developed the following jackknife procedure to obtain variance estimates of population size estimates for the homogeneous selection setup. The procedure is outlined as follows. For each $i\ \epsilon\ S_0$ define $\hat{N}_{(i)}$ to be the estimate of the population size when unit $i$ is removed from $S_0$. Define $\hat{N}_{(\cdot)}=\sum\limits_{i \epsilon S_0}\frac{\hat{N}_{(i)}}{n_0}$ where $n_0=|S_0|$. The variance estimate is
\begin{align}
\widehat{Var}_J(\hat{N})=\frac{n_0-2}{2n_0}\bigg[\sum\limits_{i \epsilon S_0}\bigg(\hat{N}_{(i)}-\hat{N}_{(\cdot)}\bigg)^2\bigg].
\end{align}

In the heterogeneous selection setup, implications result from removing a unit from the initial sample on its contribution to estimation of the size of strata the unit is external to. Hence, the following estimator is proposed. When unit $i$ is removed from $S_0$ define $\hat{N}_{k,(i)}$ to be the estimate of the size of strata $k$ and $\hat{N}_{(i)}=\sum\limits_{k=1}^K\hat{N}_{k,(i)}$. The variance estimator is
\begin{align}
\widehat{Var}_J(\hat{N}) = \sum\limits_{k=1}^K \bigg( \frac{n_{0k}-2}{2n_{0k}} \sum\limits_{i \epsilon S_{0k}} (\hat{N}_{(i)} - \hat{N})^2\bigg).
\label{strat_var}
\end{align}
Although sampling is carried out independently between strata, there may be a positive covariance of the strata size estimates. It is therefore suggested to use an approach that results in conservative confidence intervals, such as that outlined in \cite{Chao1987}, to facilitate in meeting nominal levels of coverage.

\subsection{Population Mean Estimation}

In the one-stratum case, an unbiased estimator for the population mean $\bar{z}=\sum\limits\frac{z_i}{N}$ is the initial sample mean,
\begin{align}
\bar{z}_{S_0} = \frac{\sum\limits_{i \epsilon S_0} z_i}{n_0}.
\label{one_stratum_mean}
\end{align}
An estimate for the variance of $\bar{z}_{S_0}$ is obtained by substituting the estimate of $N$ into the standard formula to give $\widehat{Var}(\bar{z}_{S_0}) = \frac{\hat{N}-n_0}{\hat{N}} \frac{s^2}{n_0}$, where $s^2$ is the sample variance of the responses from $S_0$.

In the multi-strata setup, a consistent estimator for the population mean is
\begin{align}
\bar{z}_{S_0,st} = \frac{\sum\limits_{k=1}^K \hat{N}_k \bar{z}_{S_{0k}}}{\hat{N}}
\end{align}
where $\bar{z}_{S_{0k}}$ is the mean of the responses of units selected from stratum $k$ for the initial sample. An estimate for the variance of $\bar{z}_{S_0,st}$ is obtained by substituting the estimates of $N_k$ and $N$ into the standard formula to give $\widehat{Var}(\bar{z}_{S_0,st}) = \sum\limits_{k=1}^K \bigg(\frac{\hat{N}^2_k}{\hat{N}^2}\bigg)\frac{\hat{N}_k-n_{0k}}{\hat{N}_k} \frac{s_k^2}{n_{0k}}$, where $s_k^2$ is the sample variance of the responses from $S_{0k}$.

In some cases an estimate of the proportion of individuals in a stratum can be useful. Define $p_k=\frac{N_k}{N}$ to be the population quantity to be estimated. Then $\hat{p}_k=\frac{\hat{N}_{k}}{\hat{N}}$ is a consistent estimator for this quantity. To obtain a variance estimate for this estimate the delete-one jackknife procedure is used as follows. Define $\hat{p}_{k,(i)}=\frac{\hat{N}_{k,(i)}}{\hat{N}_{(i)}}$ to be the estimate when unit $i$ is removed from the initial sample, and $\hat{p}_{k,(\cdot)}=\frac{\sum\limits_{i\epsilon S_0}\hat{p}_{k,(i)}}{n_0}$. The standard formula with the estimate of $N$ substituted into the expression is
\begin{align}
\widehat{Var}_J(\hat{p}_k)=\frac{\hat{N}-n_0}{n_0}\frac{n_0-1}{n_0}\sum\limits_{i\epsilon S_0}\bigg(\hat{p}_k-\hat{p}_{k,(\cdot)}\bigg)^2.
\end{align}

\section{Sufficiency Result}

Recall that $d_0=\{i,y_i^{k+},y_{i,j},z_i,t_i: i,j\ \epsilon\ S, k=1,...,K\}$. Define the \textit{reduced data} to be \newline
$d_R=\{i,y_i^{k+},y_{i,j},z_i,n_{0k}: i,j\ \epsilon\ S, k=1,...,K\}$, and $\underline{N}=(N_1,...,N_K)$, $\underline{\alpha}=(\alpha_1,...,\alpha_K)$.

\textbf{Theorem:} $D_R$ is sufficient for $(\underline{N}, \underline{\alpha}, \bar{z})$.

\textbf{Proof:}
\begin{align}
P(D_0=d_0)&=\prod\limits_{k=1}^K\alpha_k^{n_{0k}}(1-\alpha_k)^{(N_k-n_{0k})}\times \notag\\
&\bigg[\prod\limits_{k=1}^K\bigg[\prod\limits_{i\epsilon S_{1k}}\bigg(1-\prod\limits_{l=1}^K(1-\beta_{l,k})^{|B_{l,i}\cap S_{0l}|}\bigg)\times \prod\limits_{j \epsilon S_{0k}}\prod\limits_{l=1}^K(1-\beta_{k,l})^{|A_{j,l}\cap (U_l\setminus S_l)|}\bigg]\bigg]\notag\\
&=g(d_R,\underline{N},\underline{\alpha},\bar{z}) \times h(d_0).
\label{NFT_expression}
\end{align}
\noindent
The statistics $|B_{l,i}\cap S_{0l}|$ and $|A_{j,l}\cap (U_l\setminus S_l)|$ in Expression \ref{NFT_expression} are functions of the sampled members' time of observation, $t_i$, number of nominations to each strata, $y_i^{k+}$, and nominations within the sample, $y_{ij}$, for all $i, j\ \epsilon\ S$, which correspond with $d_0$. Therefore, by the Neyman-Factorization Theorem $D_R$ is sufficient for $(\underline{N}, \underline{\alpha}, \bar{z})$.$\ \ \ \Box$

\begin{figure}[H]
\begin{center}
\centering
\mbox{
\subfigure{\includegraphics [width=2.2in]{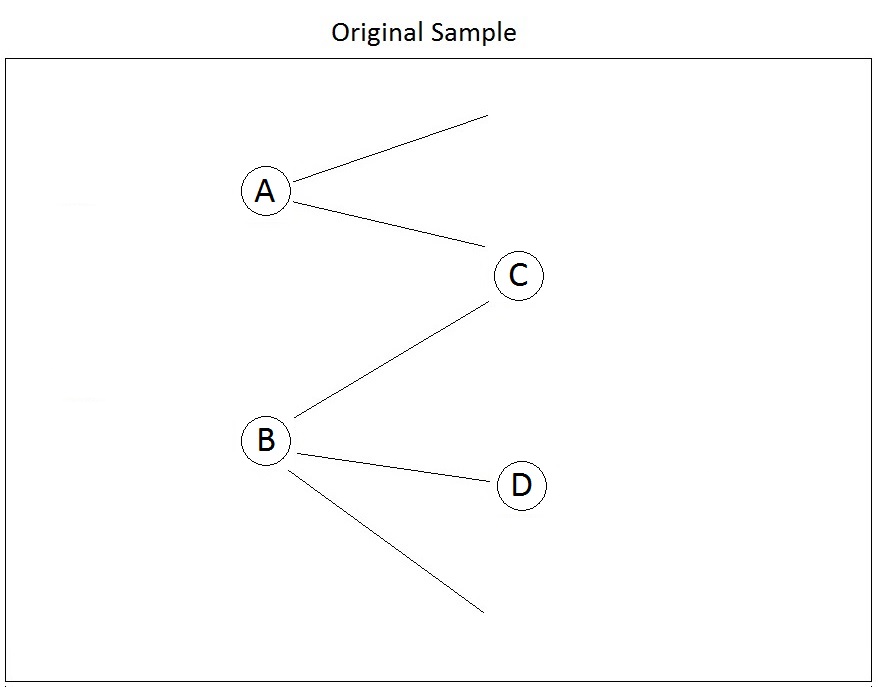}}
\subfigure{\includegraphics [width=2.2in]{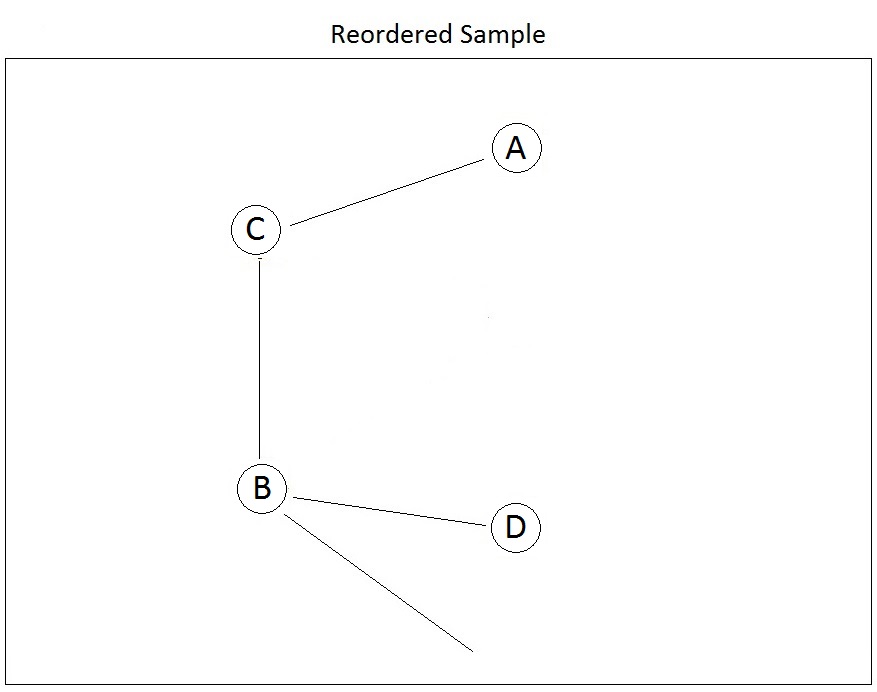}}
}
\caption{Left: Example of sample selected under sampling design outlined in Section 2. Right: Example of sample reordering consistent with reduced data of original sample.}
\label{Reorderings_example}
\end{center}
\end{figure}

Rao-Blackwellized estimates are based on evaluating selection probabilities and estimates that correspond with sample reorderings that give rise to the same reduced data. For example, under a homogenous selection setup suppose a sample is selected as presented in the left of Figure \ref{Reorderings_example}; initial sample $S_0=\{A,B\}$ is selected with probability $\alpha^2(1-\alpha)^{N-2}$, first wave $S_1=\{C,D\}$ is selected conditional on $S_0$ with probability $(1-(1-\beta)^2)\beta (1-\beta)^2$, and the corresponding estimate is $\hat{N}$, say. The reordering, labeled $v$, presented in the right of Figure \ref{Reorderings_example} is consistent with the reduced data, and $S_0^{(v)}=\{C,B\}$ is selected with probability $\alpha^2(1-\alpha)^{N-2}$, $S_1^{(v)}=\{A,D\}$ is selected conditional on $S_0^{(v)}$ with probability $\beta^2(1-\beta)$, and the corresponding estimate is $\hat{N}^{(v)}$, say.
\vspace{-5mm}
\vspace{-5mm}
There are $\mathcal{R}={n \choose n_0}$ possible sample reorderings, where $n=|S_0\cup S_1|$ and $n_0=|S_0|$. Index these as $1,...,\mathcal{R}$. The Rao-Blackwell expression is
\begin{align}
\hat{N}_{RB}=&E[\hat{N}|d_R]=\sum\limits_{v=1}^{\mathcal{R}}\bigg(\hat{N}^{(v)}P(S^{(v)}|d_R)\bigg)\notag\\
=&\frac{\sum\limits_{v=1}^{\mathcal{R}}\bigg(\hat{N}^{(v)}P(S^{(v)})\bigg)}{P(d_R)}=\frac{\sum\limits_{v=1}^{\mathcal{R}}\bigg(\hat{N}^{(v)}P(S_0^{(v)}) P(S_1^{(v)}|S_0^{(v)})\bigg)}{\sum\limits_{v=1}^\mathcal{R}\bigg(P(S_0^{(v)}) P(S_1^{(v)}|S_0^{(v)})\bigg)}
=\frac{\sum\limits_{v=1}^{\mathcal{R}}\bigg(\hat{N}^{(v)}P(S_1^{(v)}|S_0^{(v)})\bigg)}{\sum\limits_{v=1}^\mathcal{R}P(S_1^{(v)}|S_0^{(v)})}.
\label{RB_expression}
\end{align}
Note that in Expression \ref{RB_expression}, $P(S_0^{(v)})$ is constant over all $v$ and cancels from the expression, an implication of $d_R$ being sufficient for $\underline{N}$ and $\underline{\alpha}$.

Data reduction comes from mapping the set of consistent sample reorderings to the sufficient statistic, $d_R$. Hence, $d_R$ can be viewed as the set of all consistent sample reorderings and their corresponding observations. Preliminary estimation is based on the estimator corresponding with original sample ordering $d_0$, whereas Rao-Blackwellized estimates are based on a weighted average of estimates corresponding with all reorderings in $d_R$. Hence, improvement in estimation comes through utilizing more information than that provided solely with the original ordering of the sample.

\section{Markov Chain Monte Carlo}

Due to the potentially large number of reorderings, a Markov chain Monte Carlo (MCMC) procedure is used to approximate the Rao-Blackwellized estimators and their variance estimators. The procedure is outlined as follows.

Choose $M$ to be a sufficiently large number. For $m=0,1,2,...,M-1$ suppose at step $m$ of the Markov chain the most recently accepted reordering is $v$ for some $v\ \epsilon\ \{1,...,\mathcal{R}\}$. Define $p(v)$ to be the probability of selecting reordering $v$ in the full graph setting and $q(v)$ to be the probability of selecting reordering $v$ under the following proposal distribution. Define $\underline{\gamma}=(\gamma_1, \gamma_2, ..., \gamma_K)$ to be the MCMC parameters where if $K=1$ then $\gamma_1=1$, and if $K>1$ then $0 < \gamma_k < 1$ for $k=1,...,K$ such that $\sum\limits_{k=1}^K\gamma_k=1$. Sample a value $1,..., K$ with probability equal to $\gamma_k$. Suppose the sampled value is $k$. Select $k$ units from wave 1 of reordering $v$ completely at random. Interchange each of the $k$ units with one unit that nominates them from the initial sample, selected completely at random. Suppose this results in roerdering $v^*$. If the reordering is consistent with the reduced data then with probability $\text{min}\bigg\{\frac{q(v)}{q(v^*)}\frac{p(v^*)}{p(v)},1\bigg\}$ accept the proposal reordering.

The MCMC procedure starts in its stationary distribution with the original order the sample is selected in. With the aid of the $\underline{\gamma}$ parameters the chain has the potential to fully explore the distribution since pairs of units from up to $K$ different strata, where links may cross between strata, can be interchanged at any step. For example, consider the sample presented in Figure \ref{Reorderings_example}. Suppose units A and D belong to one stratum, and B and C to another. The reordering with units C and D comprising the initial sample is consistent with the reduced data, and can only be reached if units A and B are interchanged with units C and D, respectively and simultaneously. Hence, the procedure results in a Markov chain with the desired stationary distribution $P(S|d_R)$. Approximations to the Rao-Blackwellized version of a preliminary estimator and it's corresponding variance estimator based on MCMC procedures are detailed in \cite{Vincent2017}.

A test for convergence is based on the Gelman-Rubin statistic \citep{Gelman1992}. Search algorithms for two ``over-dispersed" reorderings are based on the proposal distribution, as follows. Choose $A$ to be of sufficient length for each search, and start with the original sample in the order it was selected. Suppose at some intermediate step $a=0,1,2,...,A$ the most recently accepted reordering is $v$. Draw a sample reordering, $v^*$, according to the proposal distribution. For the first over-dispersed reordering, if the reordering is consistent with the reduced data and the probability of selecting it in the full graph setting is less than that for $v$, i.e. $p(v^*) < p(v)$, then accept $v^*$. Similarly, for the second over-dispersed reordering, a consistent reordering is accepted if the probability of selecting it in the full graph setting is greater than that for $v$, i.e. $p(v^*) > p(v)$. The algorithms each conclude with their last accepted reordering, and these are used as seeds in the MCMC chains for which the convergence test is based on.

The search for the first over-dispersed reordering is likely to result in one with a corresponding smaller estimate for the population size relative to the original ordering's. The reason is that the algorithm will result in a reordering that has a smaller probability of being observed; under the sampling design this reordering will likely have more links emanating from the initial sample to individuals outside the initial sample, relative to the original ordering, because every link has a probability appended to it of being traced. Since link-tracing typically results in the selection of individuals with high-degree the search will likely result in a reordering whose initial sample is comprised of more well-connected individuals, resulting in many more nominations observed within the initial sample relative to the original ordering's. Similarly, the search algorithm for the second over-dispersed reordering is likely to result in one with a corresponding larger estimate for the population size relative to the original ordering's.

\section{Empirical Study}

\begin{figure}[H]
	\centering
\centering		
		\includegraphics[scale=0.34]{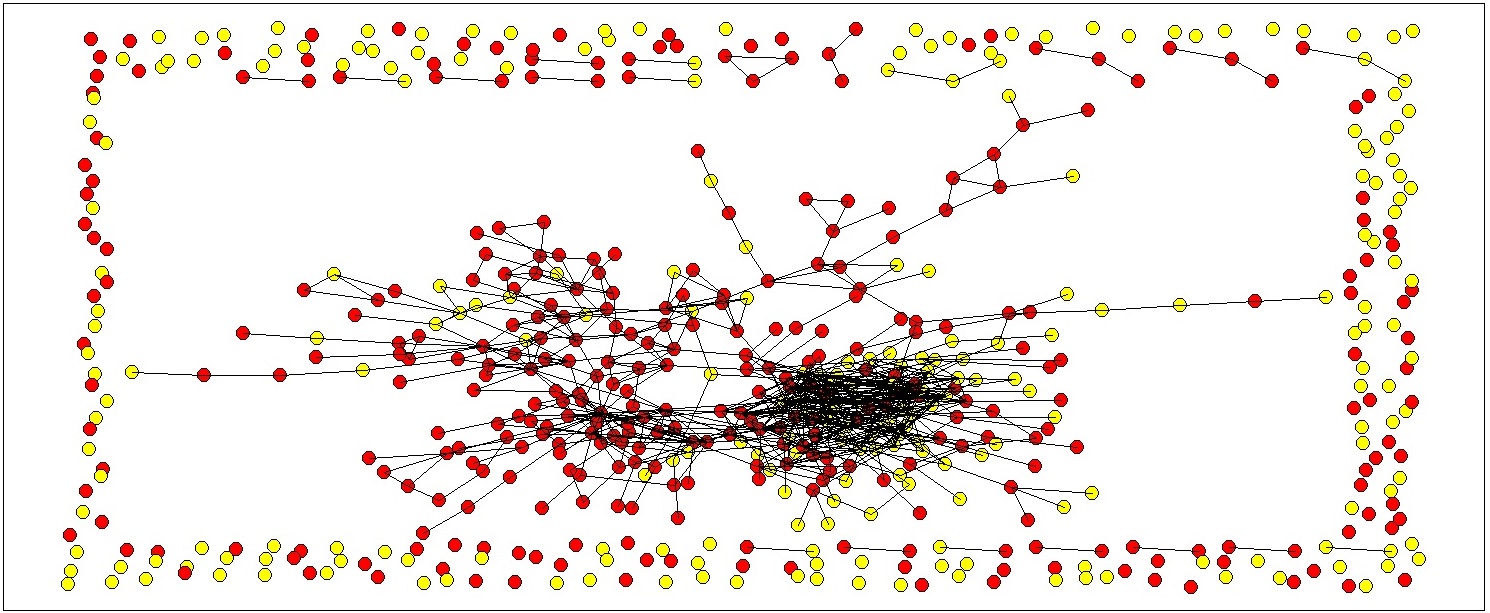}
\caption{Empirical population. Size is 595. Proportion of injection drug-users is 0.575. Average node-degree is 2.45.}	
\label{CSPop595}
\end{figure}

The empirical study is based on the P90 Colorado Springs study of 595 drug-users \citep{Darrow1999, Potterat1993, Rothenberg1995}. Figure \ref{CSPop595} gives a visual illustration of the population. The light-coloured nodes represent the stratum of non-injection drug users and dark-coloured nodes represent the stratum of injection drug-users. Links between individuals represent drug-sharing relationships. All links are reciprocated.
\vspace{-5mm}
\vspace{-5mm}
The one-stratum (homogeneous) and two-strata (heterogeneous) selection setup is considered for inference, as well as the use of a third/certainty strata; the ten individuals with largest node-degree are selected for each sample with probability one. Coverage rates and average lengths of confidence intervals corresponding with estimates for the population size are based on the log-transformation approach outlined in \cite{Chao1987}, and for the population proportion and average node-degree are based on the central limit theorem (CLT). In some cases a negative estimate is evaluated for the estimate of the variance of an improved estimate for a population quantity; there are several occurrences corresponding with an estimate for the population size with the one and two-strata setup in the second simulation study, and for the average node-degree with the three-strata setup in both simulation studies. The conservative approach presented in \cite{Vincent2017} is utilized, thereby inflating the average length of the confidence intervals.

\subsection{Simulation Study 1}

\begin{figure}[H]
	\centering
\centering		
		\includegraphics[scale=0.33]{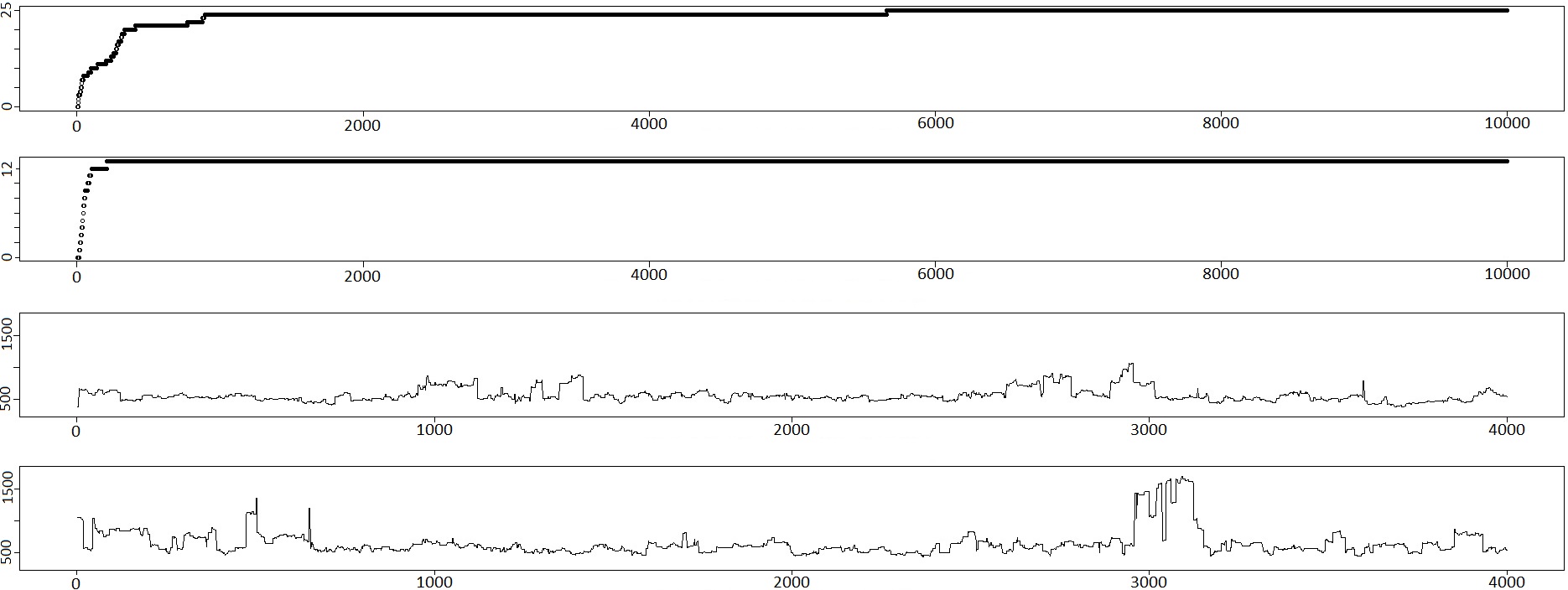}
\caption{Top: Initial sample of size $98$ represented by enlarged nodes, final sample size is $139$. Middle: Traceplots from searches for over-dispersed reorderings; change in vertical position indicates acceptance of reordering. Bottom: MCMC chains of population size estimators that start with estimates corresponding with over-dispersed reorderings; the estimates of the population size corresponding with the seed reorderings for the first and second chains are 384 and 1055, respectively. Gelman-Rubin statistic is 1.06. Preliminary population size estimate is 681, improved estimate is 618.}	
\label{Convergence_output_595}
\end{figure}

A simulation study is based on setting $\underline{\alpha}=0.15$ and $\underline{\beta}=0.20$. To determine a sufficient length of MCMC chain for approximating the Rao-Blackwellized population size estimators, the two-strata setup is considered and a search length of 10,000 is used to find over-dispersed reorderings. Figure \ref{Convergence_output_595} depicts a typical sample selected under the design with these sampling parameters, and presents output from search algorithms and MCMC chains starting with over-dispersed sample reorderings of the original sample.
\vspace{-5mm}
\begin{figure}[H]
	\centering
\centering		
		\includegraphics[scale=0.34]{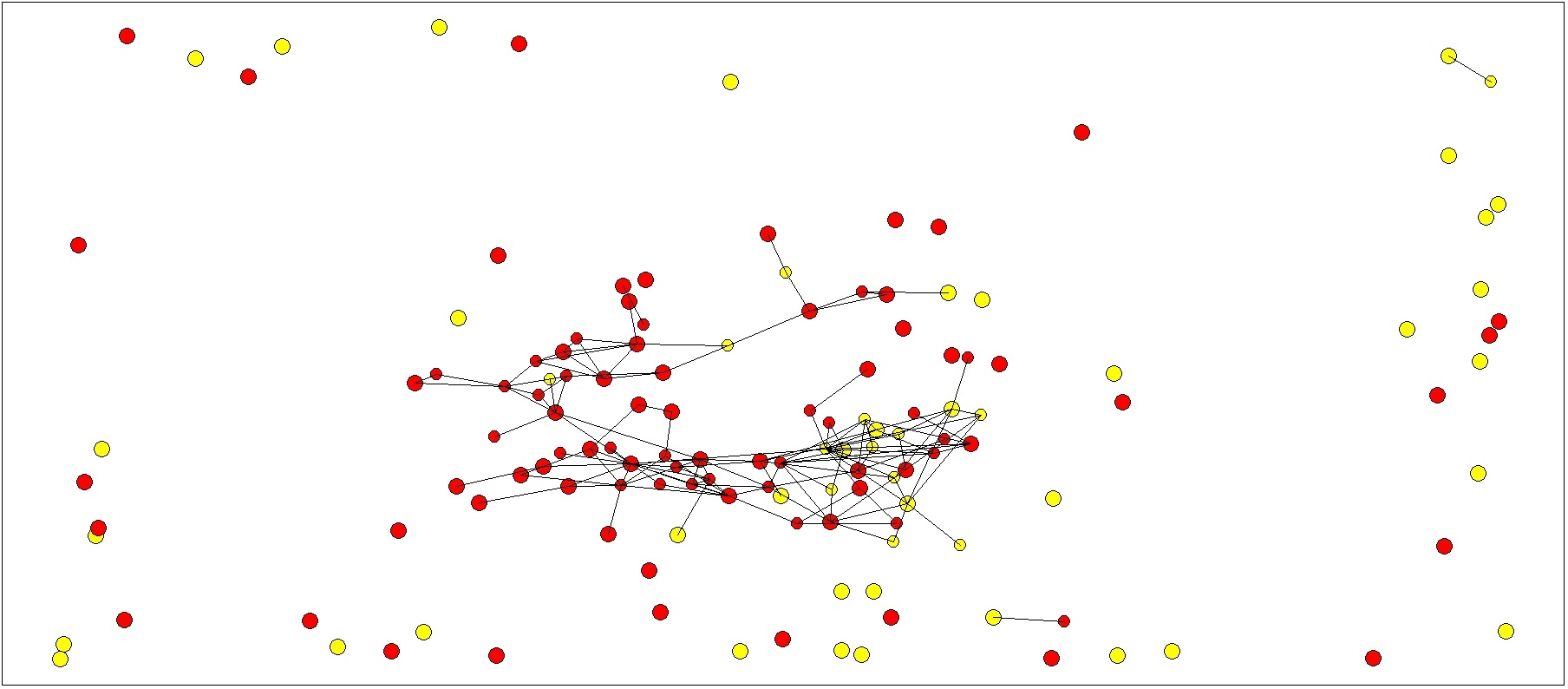}
\label{Empiricalfull}
\end{figure}
\vspace{-5mm}
\vspace{-5mm}
%Seed = 22 for output
\noindent Based on selecting 100 samples with chains set to length 2000 and $\underline{\gamma}=(0.9,0.1)$, the mean and median of the Gelman-Rubin statistics are 1.08 and 1.04, respectively.

The simulation study is based on selecting 2000 samples. The average initial sample size is 89.7 and final sample size is 122.9. Table \ref{Scores_595} presents the expectation and variance scores of the preliminary and Rao-Blackwell estimators for the population size, proportion of injection drug-users, and average node-degree. Acceptance rates for the MCMC procedure are found to be 49.6\%, 31.9\%, and 38.8\% respectively for the one-, two-, and three-strata setup. In each case a significant improvement is found with the Rao-Blackwell estimators. The use of a multi-strata setup benefits population size estimation as bias is reduced (an implication of the rate of consistency of estimators of strata size based on external nominations) and precision is increased. The estimator of the proportion of drug-users sees a decrease in precision with the two-strata setup relative to the one-stratum setup, primarily due to estimating the relative sizes of the two strata. Note that with the one-stratum setup the estimator presented in Expression \ref{one_stratum_mean} is used where indicator values are the responses. With the three-strata setup some bias is introduced since convergence rates are unequal between strata size estimators. Similarly, the estimator of the average node-degree has some bias with the two- and three-strata setup due to weighting mean responses by estimated strata sizes.
\begin{longtable}{l*{6}{r}r}
\caption{Expectation and variance scores for estimates of population size equal to 595, proportion of injection drug-users equal to 0.575, and average node-degree equal to 2.45.}
\endfirsthead
\multicolumn{6}{l}
{{Table \ref{Scores_595} continued from previous page}} \\
  \hline
Pop. Quantity       &Estimator            &Expectation           &Var. (P)           &Var. (RB)                 \\\hline
\endhead
Pop. Quantity       &Estimator            &Expectation           &Var. (P)           &Var. (RB)                 \\\hline
Size                &One-stratum          &653                   &41,796             &31,369                 \\
                    &Two-strata           &636                   &31,947             &20,230                   \\
                    &Three-strata         &617                   &20,212             &17,102                   \\\hline

Proportion          &One-stratum          &0.575                 &0.00242            &0.00212              \\
                    &Two-strata           &0.575                 &0.01130            &0.00733                \\
                    &Three-strata         &0.584                 &0.00920            &0.00755                    \\\hline

Avg. node-degree    &One-stratum          &2.446                 &0.15683            &0.12055              \\
                    &Two-strata           &2.385                 &0.16599            &0.13042                 \\
                    &Three-strata         &2.415                 &0.14349            &0.12701                 \\\hline
\label{Scores_595}
\end{longtable}
Table \ref{595coverage} provides coverage rates and average lengths of confidence intervals corresponding with the estimates. Coverage rates of the population size are close to 95\%. Hence, the pairing of the proposed variance estimator presented in Expression \ref{strat_var} and \cite{Chao1987} approach to obtaining confidence intervals are ideal for the heterogeneous setup. The average length of the confidence interval corresponding with the estimate of the population proportion under the two-strata setup is profoundly wide due to the jackknife procedure over-approximating the standard error of the estimator. Coverage rates of average node degree are less than the desired level due to substituting the estimate of strata sizes into the variance expression and skewness of the degree distribution. Further work is needed to address these issues.
\begin{longtable}{l*{6}{r}r}
\caption{Average coverage rates (CR) and length of intervals for population size estimates based on log-transformation strategy, proportion and average node-degree estimates based on CLT.}
\endfirsthead
\multicolumn{6}{l}
{{Table \ref{595coverage} continued from previous page}} \\
  \hline
Pop. Quantity       &Estimator            &CR (P)                &Length (P)         &CR (RB)               &Length (RB)        \\\hline
\endhead
Pop. Quantity       &Estimator            &CR (P)                &Length (P)         &CR (RB)               &Length (RB)        \\\hline
Size                &One-stratum          &0.970                 &848                &0.982                 &792            \\
                    &Two-strata           &0.963                 &766                &0.970                 &667             \\
                    &Three-strata         &0.967                 &632                &0.970                 &600                \\\hline

Proportion          &One-stratum          &0.942                 &0.18971            &0.945                 &0.17850          \\
                    &Two-strata           &0.959                 &0.52033            &0.987                 &0.47533           \\
                    &Three-strata         &0.959                 &0.44793            &0.964                 &0.42148                 \\\hline

Avg. node-degree    &One-stratum          &0.922                 &1.51427            &0.915                 &1.35786          \\
                    &Two-strata           &0.891                 &1.47232            &0.908                 &1.33157          \\
                    &Three-strata         &0.839                 &1.07306            &0.816                 &0.95973            \\\hline
\label{595coverage}
\end{longtable}

\subsection{Simulation Study 2}

\begin{figure}[H]
	\centering
\centering		
		\includegraphics[scale=0.33]{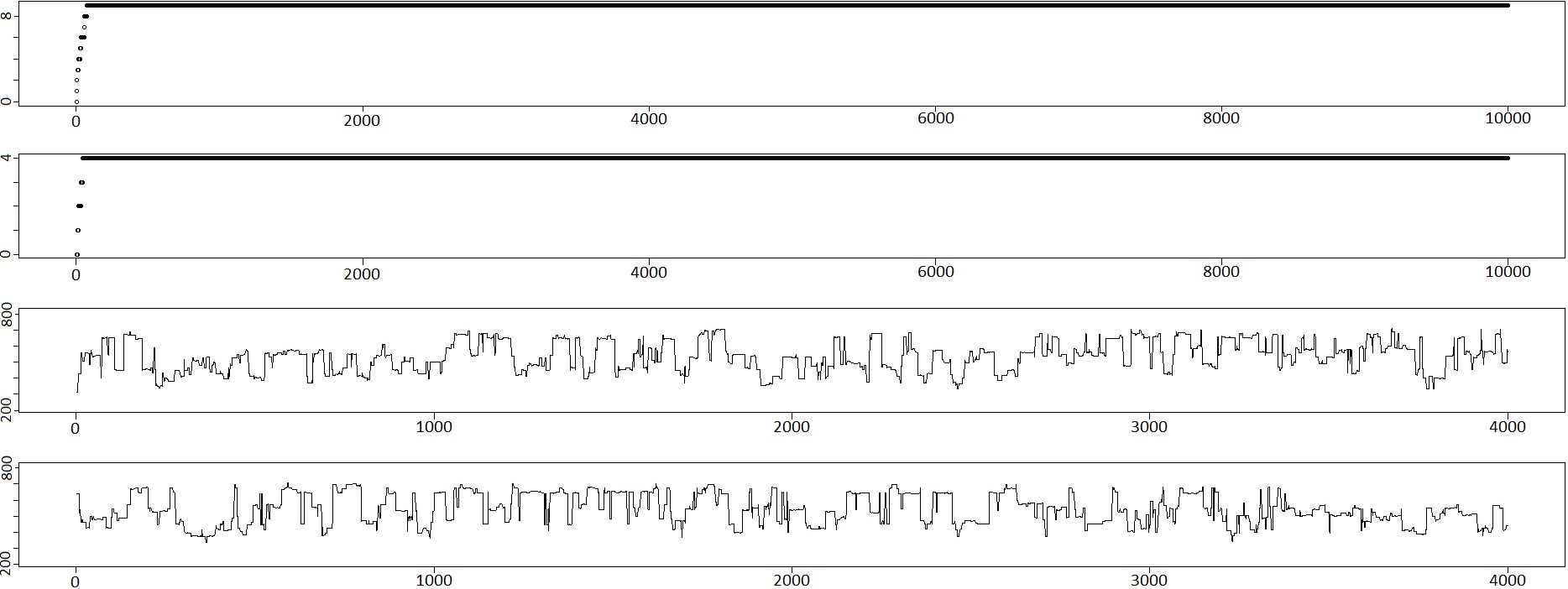}
\caption{Top: Initial sample of size $43$ represented by enlarged nodes, final sample size is $65$. Middle: Traceplots from searches for over-dispersed reorderings; change in vertical position indicates acceptance of reordering. Bottom: MCMC chains of population size estimators that start with estimates corresponding with over-dispersed reorderings; the estimates of the population size corresponding with the seed reorderings for the first and second chains are 311 and 640, respectively. Gelman-Rubin statistic is 1.02. Preliminary population size estimate is 424, improved estimate is 534.}	
\label{Convergence_output_595_2}
\end{figure}

A simulation study is based on setting $\alpha_1=0.05, \alpha_2=0.10$, and $\underline{\beta}=0.20$. To determine a sufficient length of MCMC chain for approximating the Rao-Blackwellized population size estimators, the two-strata setup is considered and a search length of 10,000 is used to find over-dispersed reorderings. Figure \ref{Convergence_output_595_2} depicts a typical sample selected under the design with these sampling parameters, where stratum one refers to the light-coloured nodes and two to the dark-coloured nodes, and presents output from search algorithms and MCMC chains starting with the original and over-dispersed sample reorderings of the original sample.
\vspace{-5mm}
\begin{figure}[H]
	\centering
\centering		
		\includegraphics[scale=0.34]{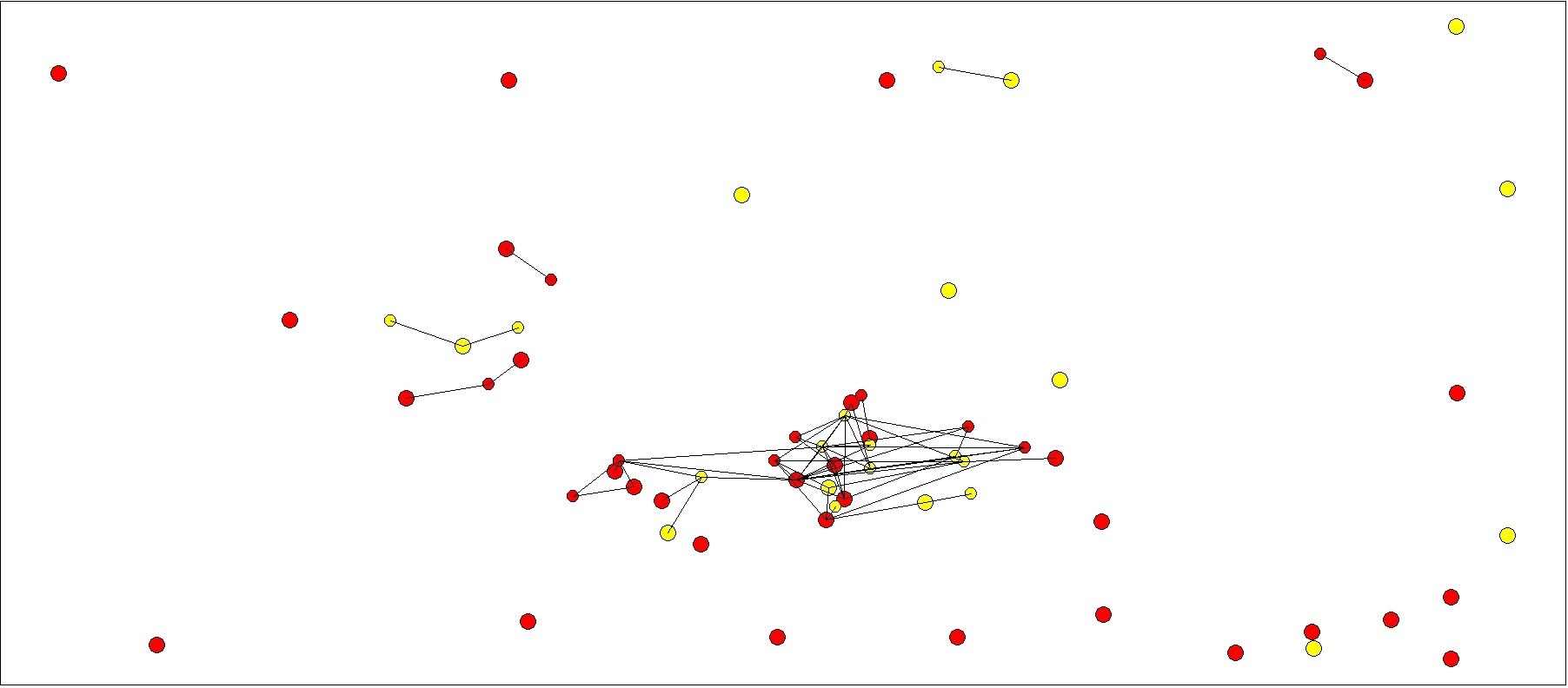}
\label{Empiricalfull_2}
\end{figure}
\vspace{-5mm}
\vspace{-5mm}
%Seed = 68 for output
\noindent Based on selecting 100 samples with chains set to length 2000 and $\underline{\gamma}=(0.9,0.1)$, the mean and median of the Gelman-Rubin statistics are 1.11 and 1.01, respectively.

The simulation study is based on selecting 2000 samples. The average initial sample size is 47.1 and final sample size is 67.5. Table \ref{Scores_595_2} presents the expectation and variance scores of the preliminary and Rao-Blackwell estimators for the population size, proportion of injection drug-users, and average node-degree. Acceptance rates for the MCMC procedure are found to be 42.7\%, 28.9\%, and 37.0\% respectively for the one-, two-, and three-strata setup. As the one stratum setup assumes homogeneity in the selection probabilities for the initial sample, the preliminary estimators and those based on the Rao-Blackwell scheme will not necessarily coincide; the expectation of the latter is given in parentheses. In this case, there seems to be close agreement between the estimators for each of the population quantities. A significant reduction in variance is seen for all Rao-Blackwell estimators. Accounting for heterogeneity in the initial sample selection procedure results in a reduction in bias for all estimators. Adding the certainty stratum helps to further reduce the bias and variance of the population size estimator.
\begin{longtable}{l*{6}{r}r}
\caption{Expectation and variance scores for estimates of population size equal to 595, proportion of injection drug-users equal to 0.575, and average node-degree equal to 2.45.}
\endfirsthead
\multicolumn{6}{l}
{{Table \ref{Scores_595_2} continued from previous page}} \\
  \hline
Pop. Quantity       &Estimator            &Expectation           &Var. (P)           &Var. (RB)                 \\\hline
\endhead
Pop. Quantity       &Estimator            &Expectation           &Var. (P)           &Var. (RB)                 \\\hline
Size                &One-stratum          &739 (729)             &409,642            &188,443                 \\
                    &Two-strata           &530                   &45,961             &29,224                   \\
                    &Three-strata         &582                   &28,541             &24,602                   \\\hline

Proportion          &One-stratum          &0.732 (0.718)         &0.00406            &0.00385              \\
                    &Two-strata           &0.672                 &0.01526            &0.01140               \\
                    &Three-strata         &0.628                 &0.01474            &0.01240                   \\\hline

Avg. node-degree    &One-stratum          &2.528 (2.530)         &0.29964            &0.26111              \\
                    &Two-strata           &2.437                 &0.29991            &0.25962                \\
                    &Three-strata         &2.436                 &0.28360            &0.25536                \\\hline
\label{Scores_595_2}
\end{longtable}
Table \ref{595coverage_2} provides coverage rates and average lengths of confidence intervals corresponding with the estimates. Due to small sample sizes and resulting bias of estimators corresponding with the one and two-sample setup the population size coverage rates are less than the desired level of 95\%. However, as seen with the three-strata setup, with enough information the variance estimator presented in Expression \ref{strat_var} and \cite{Chao1987} approach to obtaining confidence intervals are ideal. Bias in the population proportion estimates lead to coverage rates less than 95\%. Coverage rates of the average node degree are also less than 95\%, primarily due to the skewness of the distribution.
\begin{longtable}{l*{6}{r}r}
\caption{Average coverage rates (CR) and length of intervals for population size estimates based on log-transformation strategy, proportion and average node-degree estimates based on CLT.}
\endfirsthead
\multicolumn{6}{l}
{{Table \ref{595coverage_2} continued from previous page}} \\
  \hline
Pop. Quantity       &Estimator            &CR (P)                &Length (P)         &CR (RB)               &Length (RB)        \\\hline
\endhead
Pop. Quantity       &Estimator            &CR (P)                &Length (P)         &CR (RB)               &Length (RB)        \\\hline
Size                &One-stratum          &0.933                 &2737               &0.944                 &2863            \\
                    &Two-strata           &0.882                 &891                &0.906                 &799             \\
                    &Three-strata         &0.954                 &847                &0.962                 &828                \\\hline

Proportion          &One-stratum          &0.305                 &0.24291            &0.351                 &0.23399          \\
                    &Two-strata           &0.837                 &0.57700            &0.861                 &0.54061           \\
                    &Three-strata         &0.908                 &0.56555            &0.923                 &0.53933                 \\\hline

Avg. node-degree    &One-stratum          &0.927                 &2.07014            &0.912                 &1.91516          \\
                    &Two-strata           &0.905                 &2.06840            &0.907                 &1.94320        \\
                    &Three-strata         &0.797                 &1.49016            &0.796                 &1.36777          \\\hline
\label{595coverage_2}
\end{longtable}

\section{Discussion}

The new strategy is able to incorporate heterogeneity into the initial sample selection procedure, as well as to allow for individuals selected with probability one, to make significant contributions to inference. The gains in efficiency from these features are substantial. Further gains are made via Rao-Blackwellization, which directly utilizes observations of individuals sampled for the first wave.

The reduced data is sufficient for the strata sizes $\underline{N}$ and initial sample selection probabilities $\underline{\alpha}$. To implement Rao-Blackwellization, the first wave selection probabilities $\underline{\beta}$ must be known. In the empirical setting these are likely to be unknown and may have to be approximated with sample data. Alternatively, one could explore a strategy that considers the set of nominations traced from a selected individual as a random sample of fixed size, similar to the approach used in a respondent driven sampling design. Future work on these topics would be invaluable for implementing this approach in practice.

Rao-Blackwellization requires observation of the presence/absence of links between all pairs of individuals in the final sample. This may be difficult to achieve in practice. An approach that appends probabilities of links between pairs of individuals for which these are unknown, possibly based on demographic information, would make for interesting future work.

As shown in the first empirical study, it may be advantageous to base strata assignments on more than (just) initial sample selection probabilities. Further investigation into how patterns of links within and between partitions can be exploited would be worthwhile.

In some cases a complete one-wave snowball sampling design, where all links are traced, is desired and/or feasible. When this is the case it is likely that few reorderings consistent with the sufficient statistic will exist. The reason is that, as required by design, a reordering consistent with the sufficient statistic must have all units selected for the corresponding initial sample to have their links traced. Allowing sampling to continue past the first wave may permit for greater improvements in the Rao-Blackwellized estimators. For example, with a complete snowball sampling design, where link-tracing continues until there are no links out of the sample, all reorderings that retain isolated members for the initial sample will be consistent with the sufficient statistic. Furthermore, consistent reorderings under such a design will have equal probabilities of being selected in the full graph setting.

Population size estimates based directly on waves succeeding the initial sample should be explored. One approach is to base strata assignments on the distribution of links from the prior wave. For example, individuals not recently sampled and not linked to the prior wave comprise one stratum, individuals linked to one individual in the prior wave comprise another stratum, and so forth. Furthermore, if a subset of links are conceivably traced with probability one then the selected individuals comprise a certainty stratum.

%In some cases one can extend the current sufficiency result if a fixed number of nominations exist from each individual. For example, for estimating quantities of a densely connected subcomponent of a population an investigator may ask for a constant/fixed number of nominations from each sampled individual in the subcomponent. A sufficient statistic corresponding with reorderings that preserve the number of nominations within the initial sample can then be exploited for improved estimation of quantities corresponding with the subcomponent.

\section{Supplementary Materials}

The supplementary materials provide proofs required for deriving the population/strata size estimators.

%%%%%%  bibliography

%\newpage
\bibliographystyle{biom}
%\addcontentsline{Bibliography}
\bibliography{MasterReferences}

\section{Supplementary Materials}

\textbf{Claim}: For any $k=1,...,K$, $Var(r_{k,k})\leq O(\alpha_k^2N_k)$.

\textbf{Proof}: Take any $i,j\ \epsilon\ U_k$ where $i\neq j$. When squaring $r_{k,k}=\sum\limits_{\underset{a\neq b}{a,b \epsilon U_k:}} x_a x_b y_{a,b}$ this entry corresponds with
\begin{align}
&(x_{i}x_{j}y_{i,j})^2  + x_ix_jy_{i,j}x_jx_iy_{j,i} + \sum\limits_{\underset{a\neq i,j}{a\epsilon U_k:}} x_{i}x_{j}y_{i,j}x_{a}x_{j}y_{a,j} + \sum\limits_{\underset{a\neq i,j}{a\epsilon U_k:}} x_{i}x_{j}y_{i,j}x_{a}x_{i}y_{a,i} + \notag\\
&\sum\limits_{\underset{b\neq i,j}{b\epsilon U_k:}} x_{i}x_{j}y_{i,j}x_{i}x_{b}y_{i,b} + \sum\limits_{\underset{b\neq i,j}{b\epsilon U_k:}} x_{i}x_{j}y_{i,j}x_{j}x_{b}y_{j,b}
+\sum\limits_{\underset{a,b \neq i, j,\ a\neq b}{a,b \epsilon U_k:}}x_{i}x_{j}y_{i,j}x_{a}x_{b}y_{a,b}  \notag\\
&=\ast.
\end{align}

Now, the expectation of $\ast$ is
\begin{align}
E[\ast] &= \alpha_k^2 y_{i,j} + \alpha_k^2y_{i,j}y_{j,i} + \alpha_k^3y_{i,j} \sum\limits_{\underset{a\neq i,j} {a\epsilon U_k:}} y_{a,j} + \alpha_k^3y_{i,j} \sum\limits_{\underset{a\neq i,j} {a\epsilon U_k:}} y_{a,i} +
\notag\\
&\alpha_k^3y_{i,j}\sum\limits_{\underset{b\neq i,j}{b\epsilon U_k:}}y_{i,b} + \alpha_k^3y_{i,j}\sum\limits_{\underset{b\neq i,j}{b\epsilon U_k:}}y_{j,b} + \alpha_k^4y_{i,j}\sum\limits_{\underset{a,b \neq i, j,\ a\neq b}{a,b \epsilon U_k:}} y_{a,b}\notag\\
&\leq \alpha_k^2y_{i,j} + \alpha_k^2y_{i,j}y_{j,i} + 4\alpha_k^3y_{i,j}M_{k,k} + \alpha_k^4y_{i,j}(w_{k,k}-N_k).
\end{align}

Summing the above term over all $i,j\ \epsilon\ U_k$ where $i\neq j$ gives
\begin{align}
&\sum\limits_{\underset{i\neq j}{i,j \epsilon U_k:}}(\alpha_k^2y_{i,j} + \alpha_k^2y_{i,j}y_{j,i} + 4\alpha_k^3y_{i,j}M_{k,k} + \alpha_k^4y_{i,j}(w_{k,k}-N_k))\notag\\
&\leq 2\alpha_k^2(w_{k,k}-N_k) + 4\alpha_k^3M_{k,k}(w_{k,k}-N_k) + \alpha_k^4(w_{k,k}-N_k)^2.
\end{align}

Therefore,
\begin{align}
Var(r_{k,k}) &\leq 2\alpha_k^2(w_{k,k}-N_k) + 4\alpha_k^3M_{k,k}(w_{k,k}-N_k)\notag\\
&\leq 2\alpha_k^2N_kM_{k,k} + 4\alpha_k^3N_kM_{k,k}^2=\alpha_k^2N_k[2M_{k,k} + 4\alpha_kM_{k,k}^2]\notag\\
&=O(N_k\alpha_k^2).\ \ \ \Box
\end{align}

\bigskip

\textbf{Claim}: For any $k,l=1,2,...,K,$ with $k\neq l$, $Var(r_{l,k})\leq O(\alpha_l\alpha_kN_l)$.

\textbf{Proof}: Take any entry $i\ \epsilon\ U_l$ and $j\ \epsilon\ U_k$. When squaring $r_{l,k}=\sum\limits_{a \epsilon U_l, b \epsilon U_k} x_{a} x_{b} y_{a,b}$ this entry corresponds with
\begin{align}
&(x_{i}x_{j}y_{i,j})^2  + \sum\limits_{\underset{a\neq i}{a\epsilon U_l:}} x_{i}x_{j}y_{i,j}x_{a}x_{j}y_{a,j} + \sum\limits_{\underset{b\neq j}{b\epsilon U_k:}} x_{i}x_{j}y_{i,j}x_{i}x_{b}y_{i,b} + \sum\limits_{\underset{a\neq i, b\neq j}{a\epsilon U_l, b\epsilon U_k:}}x_{i}x_{j}y_{i,j}x_{a}x_{b}y_{a,b} \notag\\
&=\ast.
\end{align}

Now, the expectation of $\ast$ is
\begin{align}
E[\ast] &= \alpha_l\alpha_k y_{i,j} + \alpha_l^2\alpha_ky_{i,j} \sum\limits_{\underset{a\neq i} {a\epsilon U_l:}} y_{a,j} + \alpha_l\alpha_k^2y_{i,j}\sum\limits_{\underset{b\neq j}{b\epsilon U_k:}}y_{i,b} + \alpha_l^2\alpha_k^2y_{i,j}\sum\limits_{\underset{a\neq i, b\neq j}{a\epsilon U_l, b\epsilon U_k:}} y_{a,b}\notag\\
&\leq \alpha_l\alpha_ky_{i,j} + \alpha_l^2\alpha_ky_{i,j}M_{l,k} + \alpha_l\alpha_k^2y_{i,j}M_{l,k} + \alpha_l^2\alpha_k^2y_{i,j}w_{l,k}.
\end{align}

Summing the above term over all $i\ \epsilon\ U_l$ and $j\ \epsilon\ U_k$ gives
\begin{align}
&\sum\limits_{i \epsilon U_l, j \epsilon U_k}(\alpha_l\alpha_ky_{i,j} + \alpha_l^2\alpha_ky_{i,j}M_{l,k} + \alpha_l\alpha_k^2y_{i,j}M_{l,k} + \alpha_l^2\alpha_k^2y_{i,j}w_{l,k}) \notag\\
\leq & \alpha_l\alpha_kN_lM_{l,k} + \alpha_l^2\alpha_kN_lM_{l,k}^2+\alpha_l\alpha_k^2N_lM_{l,k}^2+\alpha_l^2\alpha_k^2w_{l,k}^2.
\end{align}

Therefore,
\begin{align}
Var(r_{l,k}) & \leq \alpha_l\alpha_kN_lM_{l,k} + \alpha_l^2\alpha_kN_lM_{l,k}^2+\alpha_l\alpha_k^2N_lM_{l,k}^2 \notag\\
& = \alpha_l\alpha_kN_l[M_{l,k}+\alpha_lM_{l,k}^2+\alpha_kM_{l,k}^2]\notag\\
& = O(\alpha_l\alpha_kN_l).\ \ \ \Box
\end{align}

\bigskip

\textbf{Claim}: For any $k=1,...,K$, $Var(s_{k,k})\leq O(\alpha_kN_k)$.

\textbf{Proof}: Take any $i,j\ \epsilon\ U_k$ where $i\neq j$. When squaring $s_{k,k}=\sum\limits_{\underset{a\neq b}{a,b \epsilon U_k:}} x_a (1-x_b) y_{a,b}$ this entry corresponds with
\begin{align}
&(x_{i}(1-x_{j})y_{i,j})^2  + \sum\limits_{\underset{a\neq i,j}{a\epsilon U_k:}} x_{i}(1-x_{j})y_{i,j}x_{a}(1-x_{j})y_{a,j} + \sum\limits_{\underset{b\neq i,j}{b\epsilon U_k:}} x_{i}(1-x_{j})y_{i,j}x_{i}(1-x_{b})y_{i,b} + \notag\\
&\sum\limits_{\underset{a,b \neq i,j,\ a\neq b}{a,b\epsilon U_k:}} x_{i}(1-x_{j})y_{i,j}x_{a}(1-x_{b})y_{a,b}=\ast.
\end{align}

Now, the expectation of $\ast$ is
\begin{align}
E[\ast] &= \alpha_k(1-\alpha_k) y_{i,j} + \alpha_k^2(1-\alpha_k)y_{i,j} \sum\limits_{\underset{a\neq i,j} {a\epsilon U_k:}} y_{a,j} + \alpha_k(1-\alpha_k)^2y_{i,j}\sum\limits_{\underset{b\neq i,j}{b\epsilon U_k:}}y_{i,b} + \notag\\
 &\ \ \ \ \ \ \alpha_k^2(1-\alpha_k)^2y_{i,j}\sum\limits_{\underset{a,b\neq i,j,\ a\neq b} {a,b\epsilon U_k:}} y_{a,b}\notag\\
&\leq \alpha_k(1-\alpha_k)y_{i,j} + \alpha_k^2(1-\alpha_k)y_{i,j}M_{k,k} + \alpha_k(1-\alpha_k)^2y_{i,j}M_{k,k}+\notag\\
&\ \ \ \ \ \ \alpha_k^2(1-\alpha_k)^2y_{i,j}(w_{k,k}-N_k).
\end{align}

Summing the above term over all $i,j\ \epsilon\ U_k$ where $i \neq j$ gives
\begin{align}
&\sum\limits_{\underset{i\neq j}{i,j \epsilon U_k:}}(\alpha_k(1-\alpha_k)y_{i,j} + \alpha_k^2(1-\alpha_k)y_{i,j}M_{k,k} + \alpha_k(1-\alpha_k)^2y_{i,j}M_{k,k}+\notag \\
&\ \ \ \ \ \ \alpha_k^2(1-\alpha_k)^2y_{i,j}(w_{k,k}-N_k))\notag\\
\leq & \alpha_k(1-\alpha_k)N_kM_{k,k} + \alpha_k^2(1-\alpha_k)N_kM_{k,k}^2 + \alpha_k(1-\alpha_k)^2N_kM_{k,k}^2 + \alpha_k^2(1-\alpha_k)^2(w_{k,k}-N_k)^2.
\end{align}

Therefore,
\begin{align}
Var(s_{k,k}) &\leq \alpha_k(1-\alpha_k)N_kM_{k,k} + \alpha_k^2(1-\alpha_k)N_kM_{k,k}^2 + \alpha_k(1-\alpha_k)^2N_kM_{k,k}^2\notag\\
&=\alpha_kN_k[(1-\alpha_k)M_{k,k}+\alpha_k(1-\alpha_k)M_{k,k}^2+(1-\alpha_k)^2M_{k,k}^2]\notag\\
&=O(\alpha_kN_k).\ \ \ \Box
\end{align}

\bigskip

\textbf{Claim}: For any $k,l=1,2,...,K,$ with $k\neq l$, $Var(s_{l,k})\leq O(\alpha_lN_l)$.

\textbf{Proof}: Take any entry $i\ \epsilon\ U_l$ and $j\ \epsilon\ U_k$. When squaring $s_{l,k}=\sum\limits_{a \epsilon U_l, b \epsilon U_k} x_{a} (1-x_{b}) y_{a,b}$ this entry corresponds with
\begin{align}
&(x_{i}(1-x_{j})y_{i,j})^2  + \sum\limits_{\underset{a\neq i}{a\epsilon U_l:}} x_{i}(1-x_{j})y_{i,j}x_{a}(1-x_{j})y_{a,j} + \sum\limits_{\underset{b\neq j}{b\epsilon U_k:}} x_{i}(1-x_{j})y_{i,j}x_{i}(1-x_{b})y_{i,b} + \notag\\
&\sum\limits_{\underset{a\neq i, b\neq j}{a\epsilon U_l, b\epsilon U_k:}}x_{i}(1-x_{j})y_{i,j}x_{a}(1-x_{b})y_{a,b}=\ast.
\end{align}

Now, the expectation of $\ast$ is
\begin{align}
E[\ast] &= \alpha_l(1-\alpha_k) y_{i,j} + \alpha_l^2(1-\alpha_k)y_{i,j} \sum\limits_{\underset{a\neq i} {a\epsilon U_l:}} y_{a,j} + \alpha_l(1-\alpha_k)^2y_{i,j}\sum\limits_{\underset{b\neq j}{b\epsilon U_k:}}y_{i,b} +\notag\\
&\ \ \ \ \ \ \alpha_l^2(1-\alpha_k)^2y_{i,j}\sum\limits_{\underset{a\neq i, b\neq j}{a\epsilon U_l, b\epsilon U_k:}}y_{a,b}\notag\\
&\leq \alpha_l(1-\alpha_k)y_{i,j} + \alpha_l^2(1-\alpha_k)y_{i,j}M_{l,k} + \alpha_l(1-\alpha_k)^2y_{i,j}M_{l,k} + \alpha_l^2(1-\alpha_k)^2y_{i,j}w_{l,k}.
\end{align}

Summing the above term over all $i\ \epsilon\ U_l$ and $j\ \epsilon\ U_k$ gives
\begin{align}
&\sum\limits_{i \epsilon U_l, j \epsilon U_k}(\alpha_l(1-\alpha_k)y_{i,j} + \alpha_l^2(1-\alpha_k)y_{i,j}M_{l,k} + \alpha_l(1-\alpha_k)^2y_{i,j}M_{l,k} + \alpha_l^2(1-\alpha_k)^2y_{i,j}w_{l,k}) \notag\\
\leq & \alpha_l(1-\alpha_k)N_lM_{l,k} + \alpha_l^2(1-\alpha_k)N_lM_{l,k}^2+\alpha_l(1-\alpha_k)^2N_lM_{l,k}^2+\alpha_l^2(1-\alpha_k)^2w_{l,k}^2.
\end{align}

Therefore,
\begin{align}
Var(s_{l,k}) & \leq \alpha_l(1-\alpha_k)N_lM_{l,k} + \alpha_l^2(1-\alpha_k)N_lM_{l,k}^2+\alpha_l(1-\alpha_k)^2N_lM_{l,k}^2 \notag\\
& = \alpha_lN_l[(1-\alpha_k)M_{l,k}+\alpha_l(1-\alpha_k)M_{l,k}^2+(1-\alpha_k)^2M_{l,k}^2]\notag\\
& = O(\alpha_lN_l). \ \ \ \Box
\end{align}

\end{document}